\documentclass{proc}

\begin{document}

\title{Warped super--bigravity}

\author{ZYGMUNT LALAK AND RADOS\L AW MATYSZKIEWICZ}

\address{Institute of Theoretical Physics, University of Warsaw, Poland\\ 
E-mail: Zygmunt.Lalak@fuw.edu.pl, rmatysz@fuw.edu.pl}


\begin{flushright} 
IFT-2003-32
\end{flushright}
\vspace{.4em}
\begin{center}{\bf WARPED SUPER--BIGRAVITY}\end{center}
\vspace{1.3em}
\begin{center}ZYGMUNT LALAK AND RADOS\L AW MATYSZKIEWICZ\end{center}
\begin{center}{\it Institute of Theoretical Physics, University of Warsaw, Poland}\\ 
{\it E-mail: Zygmunt.Lalak@fuw.edu.pl, rmatysz@fuw.edu.pl}\end{center}

\vspace{.8em}
\abstracts{Warped supergravities with lower--dimensional branes 
provide consistent field--theoretic framework 
for discussion of important aspects of modern string theory physics.
In particular, it is possible to study reliably supersymmetry breakdown and its transmission between branes. In this talk we present two versions of locally 
supersymmetrized bigravity -- one with unbroken $N=1, \, d=4$ supersymmetry, 
and the second one with all supersymmetries broken by 
boundary conditions.}

\section{Introduction}
Brane-bulk supergravities generalize the concept of supersymmetry to 
setups combining degrees of freedom  that propagate in subspaces of various dimensionalities, sometimes spatially disconnected \cite{Altendorfer:2000rr,Falkowski:2000er}. 
They provide  consistent field--theoretic framework 
for discussion of important aspects of modern string theory physics.
In particular, with supergravities that include lower dimensional branes it 
is possible to study reliably supersymmetry breakdown and its transmission between different sectors of a model. 
Sources localized on lower--dimensional branes provide contributions
to the energy--momentum tensor, and result in the appearance of the warp 
factor that multiplies the four-dimensional metric. In this note we shall 
discuss a special class of warped supergravities where compactification 
gives (AdS--)bigravity in four dimensions. It is interesting to 
observe that supersymmetry implies the existence of ultra-light 
gravitini. We discuss spectra of gravitons and gravitini obtained upon 
KK decomposition of five--dimensional fields and localization of 
corresponding wave functions. We shall demonstrate how supersymmetry 
breaking due to twisting of boundary conditions for gravitini influences 
 the spectra, and shapes of the wave functions.  

\section{General setup}
To begin with, let us briefly summarize the brane--bulk
super--bigravity Lagrangian, constructed in \cite{Lalak:2003fu}. The simple N=2 d=5 supergravity multiplet contains metric tensor (represented by 
the vielbein $e^m_\alpha$), two gravitini $\Psi^A_\alpha$ and one vector field $A_\alpha$ -- the graviphoton. We shall consider gauging of a $U(1)$ subgroup of the global $SU(2)_R$ symmetry of the 5d Lagrangian. In general, coupling of bulk fields to branes turns out to be related to the gauging, and the bulk--brane couplings will preserve only a subgroup of the $SU(2)_R$. We shall add to the initial bulk Lagrangian boundary terms that include brane tension and/or gravitini mass term on the brane. The
5d action describing such a setup reads 
        $S=\int_{M_5}\ e_5 \ ({\cal L}_{bulk}+{\cal L}_{brane})$,
         where
       	\begin{equation} 
        {\cal L}_{brane} = \sum_i e_4\delta(y-y_i)\left(-\lambda_i- \bar{\Psi}_\mu^A\gamma^{\mu\nu}(M_i+\gamma_5\bar{M_i})_{A}^{\;B} \Psi_{\nu B}\right) \ .
   \end{equation} 
        The $M_i$, $\bar{M_i}$ are constant and symmetric matrices that denote gravitini mass terms on the branes at the fixed points $y_i$. Covariant derivative contains both gravitational and gauge connections:
        $
        D_\alpha\Psi_\beta^A=\nabla_\alpha\Psi_\beta^A+ A_\alpha{\cal P}^A_{\;B}\Psi_\beta^B
        $,  
where ${\cal P}={\cal P}_a\,{\rm i}\sigma^a$ is the gauge prepotential. The pair of gravitini satisfies symplectic Majorana\- condition
 $\bar{\Psi}^A\equiv\Psi_A^\dagger\gamma_0=(\epsilon^{AB}\Psi_B)^TC$ where $C$ is the charge conjugation matrix and $\epsilon^{AB}$ is antisymmetric $SU(2)_R$ metric (we use the convention $\epsilon_{12}=\epsilon^{12}=1$). 
Supersymmetry transformations include singular terms proportional to the delta functions
        \begin{eqnarray} 
       &&\delta\Psi_{\alpha}^A=D_\alpha\eta^A- \frac{{\rm i}}{4\sqrt{2}}\left(\gamma_\alpha^{\;\beta\gamma}-4\delta_\alpha^{\;\beta}\gamma^\gamma\right){\cal F}_{\beta\gamma}\eta^A+\frac{\sqrt{2}{\rm i}}{3}{\cal P}^{AB}\gamma_\alpha\eta_B\ \nonumber\\&&\qquad\; + \epsilon^{-1}(y)\delta_\alpha^{\;5}\sum_ia_i\delta(y-y_i)(Q_i-\gamma_5\delta)^{A}_{\;B}\gamma_5\eta^B,
        \end{eqnarray} 
	where $a_i=1$ if $\epsilon(y)$ 'jumps up' at the fixed point $y_i$, or $a_i= -1$ if it 'jumps down'. The $Q_i$'s denote ${\bf Z}_2$ operators acting in the gravitino sector as follows
	\begin{equation} \label{gbcond}
        \Psi^A_\mu(y_i-y)=\gamma_5(Q_i)^A_{\;B}\Psi^B_\mu(y_i+y)\ ,\quad\Psi^A_5(y_i-y)=-\gamma_5(Q_i)^A_{\;B}\Psi^B_5(y_i+y)\ .
	\end{equation}
The symplectic Majorana condition 
and the normalization $(Q_i)^2=1$ imply $Q_i=(q_i)_a \sigma^a$, where $(q_i)_a$ are real parameters \cite{Bergshoeff:2000zn}.

In the general case \cite{Brax:2001xf} one  can write down the prepotential in the following way:
     $ P = g_R \epsilon(y) R + g_{S} S$,
     where $R=r_a\, {\rm i} \sigma^a$ commutes and  $S=s_a\, {\rm i}\sigma^a$ anticommutes with each $Q_i$.

  The closure of the supersymmetry algebra provides the relations  between parameters of the
boundary Lagrangian and the prepotential 
\cite{Bagger:2002rw,Lalak:2003fu} :
  \begin{eqnarray}
    0=&&\delta(y-y_i)\bar{\Psi}_\mu^A\gamma^\mu\bigg[(M_i-\gamma_5\bar{M}_i)_{A}^{\;B}P_{B}^{\;C}+a_i\frac{1}{2}\epsilon^{-1}(y)P_{A}^{\;B}(Q_i+\gamma_5\delta)_{B}^{\;C}\nonumber\\&&+a_ig_R\gamma_5 R_{A}^{\;C}+\frac{{\rm i}}{4\sqrt{2}}\lambda_i\delta_{A}^{\;C}+a_i\frac{{\rm i}}{4\sqrt{2}}\lambda_i\epsilon(y)\left(\gamma_5(M_i)_{A}^{\;C}-(\bar{M}_i)_{A}^{\;C}\right)\nonumber\\&&+\frac{{\rm i}}{4\sqrt{2}}\lambda_i\left(\frac{1}{2}\gamma_5(Q_i)_{A}^{\;C}+\frac{1}{2}\delta_{A}^{\;C}\right)\bigg]\eta_C\ .
  \end{eqnarray}   

\section{Super--bigravity}
Let us construct 5d supergravity with two positive tension branes at the points $y=0$ and $y=\pi r_c$. This model has interesting cosmological implications, especially for a large fifth dimension (bigravity proposed in \cite{Kogan:1999wc,Kogan:2000vb}). To simplify the discussion let us assume the prepotential of the form: $P_A^{\;B}=g{\rm i}(\sigma_1)_A^{\;B}$ and $(Q_0)_A^{\;B}=(Q_\pi)_A^{\;B}=(\sigma_3)_A^{\;B}$. 
Gravitini masses on the branes read
     \begin{equation}
	 (M_{0,\pi})_A^{\;B}=\frac{1}{2}\alpha_{0,\pi}(\sigma_1)_A^{\;B}\ ,\quad (\bar{M}_{0,\pi})_A^{\;B}=\frac{1}{2}\alpha_{0,\pi}{\rm i}
(\sigma_2)_A^{\;B}\ .
       \end{equation}
Taking
      \begin{equation} \label{alphamasy} 
	\alpha_0=-\frac{\cosh(k\pi r_{c}/2)\pm 1}{\sinh(k\pi r_{c}/2)}\ ,\quad	\alpha_\pi=-\frac{\cosh(k\pi r_{c}/2)\pm 1}{\sinh(k\pi r_{c}/2)},
      \end{equation} 
      we obtain bosonic action of the bigravity model:
       \begin{equation} \label{lagads}
	S= \int d^5 x \sqrt{-g_5} (\frac{1}{2}R + 6 k^2)-  6 \int d^5 x\sqrt{-g_4}k T (\delta(y) +\delta(y-\pi r_c))\ ,
       \end{equation}
       where $k=\frac{2\sqrt{2}}{3}g_R$ and $T=\tanh(k\pi r_{c}/2)$.
       Gravitational background does not admit a flat 4d Minkowski foliation, and the consistent 
solution is that of $AdS_4$ branes:
        \begin{equation} 
        ds^{2}=a^{2}(y)\bar{g}_{\mu\nu}dx^{\mu}dx^{\nu}+dy^{2}\ ,
        \end{equation}  
        where 
        \begin{equation} 
        a(y)=\frac{\sqrt{-\bar{\Lambda}}}{k}\cosh\left(k|y|-\frac{k\pi r_{c}}{2}\right)\ ,
        \end{equation} 
        and $\bar{g}_{\mu\nu}dx^{\mu}dx^{\nu}=\exp(-2\sqrt{-\bar{\Lambda}}x_{3})(-dt^{2}+dx^{2}_{1}+dx_{2}^{2})+dx_{3}^{2}$ is the four dimensional $AdS$ metric. 

The radius of the fifth dimension is determined in terms of the brane tensions
        \begin{equation} 
        k\pi r_{c}=\ln\left(\frac{1+T}{1-T}\right)\ .
        \end{equation} 
        Normalization $a(0)=1$ leads to the fine tuning relation $\bar{\Lambda}=(T^{2}-1)k^{2}<0$.
       
 Notice in (\ref{alphamasy}), that we have two possibilities for the brane 
gravitini masses: $\alpha_0=1/\alpha_\pi$ and $\alpha_0=\alpha_\pi$. One can check that the first one is directly related to the BFL model \cite{Brax:2001xf}
 of twisted supergravity (compactification of this model has been performed 
in \cite{Lalak:2002kx}). In the first case five--dimensional vacuum spontaneously brakes all supersymmetries), hence we shall start with the detailed exploration  of the other one. For simplicity, let us assume 
$\alpha_0=\alpha_\pi=-\alpha$, where
        \begin{equation} 
	\alpha=\frac{\cosh(k\pi r_{c}/2)- 1}{\sinh(k\pi r_{c}/2)}\ .
      \end{equation} 
      Then the boundary conditions take the form 
         \begin{eqnarray} 
	&&\epsilon^{-1}(y) \delta(y) \gamma_5(\Psi^-_\mu)^A=\delta(y)\alpha(\sigma_1)^A_{\;B}(\Psi^+_\mu)^B\ ,\label{warbrzadsbigravit1}\\	&&\epsilon^{-1}(y) \delta(y-\pi r_c) \gamma_5(\Psi^-_\mu)^A=-\delta(y-\pi r_c)\alpha(\sigma_1)^A_{\;B}(\Psi^+_\mu)^B\label{warbrzadsbigravit2}\ ,
      \end{eqnarray} 
	 where we have defined $(\Psi^\pm_{\nu})^{A}=\frac{1}{2}(\delta\pm\gamma_5\sigma_3)^A_{\;B}\Psi_{\nu}^{B}$.

      One can show (see \cite{Lalak:2003fu}) that supersymmetry is preserved in the effective theory and the Killing spinor reads:
         \begin{eqnarray} \label{killingwector}
	    &&\eta^A=N\cosh\left(k|y|/2-k\pi r_{c}/4\right)\left(\begin{array}{c}\hat{\eta}_R\\-\hat{\eta}_L\end{array}\right)^A\nonumber\\&&\quad\;-N\epsilon(y)\sinh\left(k|y|/2-k\pi r_{c}/4\right)\left(\begin{array}{c}\hat{\eta}_L\\\hat{\eta}_R\end{array}\right)^A\ ,
      \end{eqnarray} 
	  where $\hat{\eta}$ denotes Killing spinor in the $AdS_4$, which satisfies: 
      $\left(\bar{\nabla}_\mu-\frac{1}{2}\sqrt{-\bar{\Lambda}}\hat{\gamma}_\mu\right)\hat{\eta}=0$,
      and $N$ is a normalization constant.

       KK modes of the graviton have been found in \cite{Lalak:2002kx}. Let us briefly quote the results. One can write small fluctuations around vacuum metric as $g_{\mu\nu}(x^{\rho},y)=a^{2}(y)\bar{g}_{\mu\nu}+\phi_{h}(y)h_{\mu\nu}(x^{\rho})$, where $h_{\mu\nu}(x^{\rho})$ is a 4d wave function in $AdS_{4}$  background ($(\nabla_\rho\nabla^\rho+ 2\bar{\Lambda})h_{\mu\nu}=m^{2}h_{\mu\nu}$). The spectrum contains the massless mode $\phi_h=A_{0}\cosh^{2}(k|y|-k\pi r_{c}/2)$ and the tower of massive modes:    
        \begin{eqnarray}
        \phi_h=&A_{m}{\rm LP}\left(\frac{1}{2}\left(-1+\sqrt{1+4\bar{m}^{2}}\right),2,\tanh\left(k|y|-\frac{k\pi r_{c}}{2}\right)\right)&\nonumber\\&+B_{m}{\rm LQ}\left(\frac{1}{2}\left(-1+\sqrt{1+4\bar{m}^{2}}\right),2,\tanh\left(k|y|-\frac{k\pi r_{c}}{2}\right)\right)&\ ,
        \end{eqnarray}  
        where ${\rm LP}(m,n,x)$ and ${\rm LQ}(m,n,x)$ are associated Legendre functions of the first and second kind respectively. We have introduced the new symbol $\bar{m}=\sqrt{-m^{2}/\bar{\Lambda}+2}$. Matching delta functions at fixed points leads to the  following mass quantization condition
        \begin{eqnarray} \label{warunkinamasyadsgrawiton}
        0=&\left(2t{\rm LQ}(,,-t) +c{\rm LQ}'(,,-t)\right)\left(-2t{\rm LP}(,,t) +c{\rm LP}'(,,t)\right)+&\nonumber\\&-\left(2t{\rm LP}(,,-t)+ c{\rm LP}'(,,-t)\right)\left(-2t{\rm LQ}(,,t)+c{\rm LQ}'(,,t)\right)&,
        \end{eqnarray}  
        where we have introduced notation $t=\tanh(k\pi r_{c}/2)$ and $c=\cosh^{-2}(k\pi r_{c}/2)$.

    Upon the factorization
        \begin{eqnarray} \label{decompgravitini}
        &&(\Psi^{+}_{\mu})^A=\phi^{+}_1(y)\left(\begin{array}{c}(\hat{\psi}^+_{\mu})_R\\-(\hat{\psi}^+_\mu)_L\end{array}\right)^A+\phi^{-}_2(y)\left(\begin{array}{c}(\hat{\psi}^-_{\mu})_R\\-(\hat{\psi}^-_\mu)_L\end{array}\right)^A\ ,\nonumber\\ &&(\Psi^{-}_{\mu})^A=\epsilon(y)\phi^{+}_2(y)\left(\begin{array}{c}(\hat{\psi}^+_{\mu})_L\\(\hat{\psi}^+_\mu)_R\end{array}\right)^A+\epsilon(y)\phi^{-}_1(y)\left(\begin{array}{c}(\hat{\psi}^-_{\mu})_L\\(\hat{\psi}^-_\mu)_R\end{array}\right)^A\ ,
        \end{eqnarray} 
	where $\hat{\psi}^\pm_\mu$ denote 4d gravitini in $AdS_4$:
       $
  \gamma^{\mu\rho\nu}\nabla_{\rho}\hat{\psi}^\pm_{\nu}=\mp\sqrt{m_\pm^{2}-\bar{\Lambda}}\;\gamma^{\mu\nu}\hat{\psi}^\pm_{\nu},
        $
        the equations of motion for the even and odd components of gravitini 
        lead to the equations
        \begin{eqnarray} \label{equationgravitads}
        &&\phi^{\pm\prime}_1+k\epsilon(y)\tanh\left(k|y|-\frac{k\pi r_{c}}{2}\right)\phi^{\pm}_1+\frac{3}{2}\phi^{\pm}_{2}+ a^{-1}(y)\sqrt{m_\pm^{2}-\bar{\Lambda}}\;\phi^{\pm}_{2}=0\ ,\nonumber\\ &&\phi^{\pm\prime}_2+k\epsilon(y)\tanh\left(k|y|-\frac{k\pi r_{c}}{2}\right)\phi^{\pm}_2+\frac{3}{2}\phi^{\pm}_{1}- a^{-1}(y)\sqrt{m_\pm^{2}-\bar{\Lambda}}\;\phi^{\pm}_{1}=0\ .
        \end{eqnarray}
	and to boundary conditions      
        \begin{eqnarray} \label{wargravitinoads}
        &&\phi_{2}^{+}(0)=\tanh\left(\frac{k\pi r_{c}}{4}\right)\phi_{1}^{+}(0)\ ,\quad  \phi_{2}^{+}(\pi r_c)=-\tanh\left(\frac{k\pi r_{c}}{4}\right)\phi_{1}^{+}(\pi r_c)\ ,\nonumber\\ &&\phi_{2}^{-}(0)=\coth\left(\frac{k\pi r_{c}}{4}\right)\phi_{1}^{-}(0)\ ,\quad  \phi_{2}^{-}(\pi r_c)=-\coth\left(\frac{k\pi r_{c}}{4}\right)\phi_{1}^{-}(\pi r_c)\ .
        \end{eqnarray} 
	These constraints remove a massless mode from the spectrum of the gravitino $\hat{\psi}^-_\mu$. However, for $\hat{\psi}^+_\mu$ the massless mode exists, and the solutions of the equation (\ref{equationgravitads}) can be written down as follows
	 \begin{equation} 
	    \phi^+_1(y)= N\cosh\left(k|y|/2-k\pi r_{c}/4\right)\ ,\quad \phi^+_2(y)=-N\sinh\left(k|y|/2-k\pi r_{c}/4\right)\ .
	 \end{equation} 
	 This is what we have expected, because only a half of the 
supersymmetry generators is preserved in the effective 4d theory. One can also find  massive modes. They are expressible through hypergeometric functions. 
The condition (\ref{wargravitinoads}) leads to the mass quantization.
      Let us recall that the $AdS_4$ supermultiplets contain in general particles with different mass terms. For instance, massive higher spin representations ($E_0 
> s+1, \, s \geq 1/2$) are of the form 
\begin{equation}
D(E_0,s) \oplus D(E_0 +1/2, s+ 1/2) \oplus D(E_0 + 1/2, s-1/2) \oplus D(E_0 +1, s),
\end{equation} 
with the mass-squared operator $m^2= E_0 (E_0 -3) - (s+1)(s-2)$. This implies the 
spin-2 and spin-3/2 spectra 
    \begin{eqnarray} \label{spectads}
      && m^{2}_{2} = (E_0 + 1/2)(E_0 -5/2)\ ,\nonumber\\
      && m^{2}_{3/2^+}= E_0(E_0 -3) +5/4\ ,\nonumber\\ 
      && {m}^{2}_{3/2^-}= (E_0 +1)(E_0 -2) +5/4\ ,
      \end{eqnarray}
for some $E_0$ (in units of $\sqrt{-\bar{\Lambda}}$). 
  One can check that in the limit $k r_c\ll 1$ masses are quantized in units of
 $1/r_{c}$ and obey the relations (\ref{spectads}).    
      Notice that spectra of gravitini are shifted with respect to these of the graviton by $\pm \frac{1}{2}\sqrt{-\bar{\Lambda}}$. It means that the shift 
between the two gravitini towers is $\sqrt{-\bar{\Lambda}}$. 

\begin{figure}[h]
        \begin{center}
        \epsfig{file=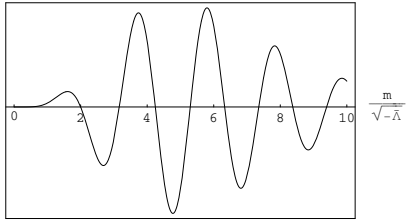, width=.45 \linewidth}
        \caption{\small{The zeros corresponds to the mass spectrum of the graviton,
($k r_{c}\pi=10$).} \label{masygravitonu10}}  
        \end{center}
\end{figure}

\begin{figure}[h]    
	\begin{center}                                        
        \epsfig{file=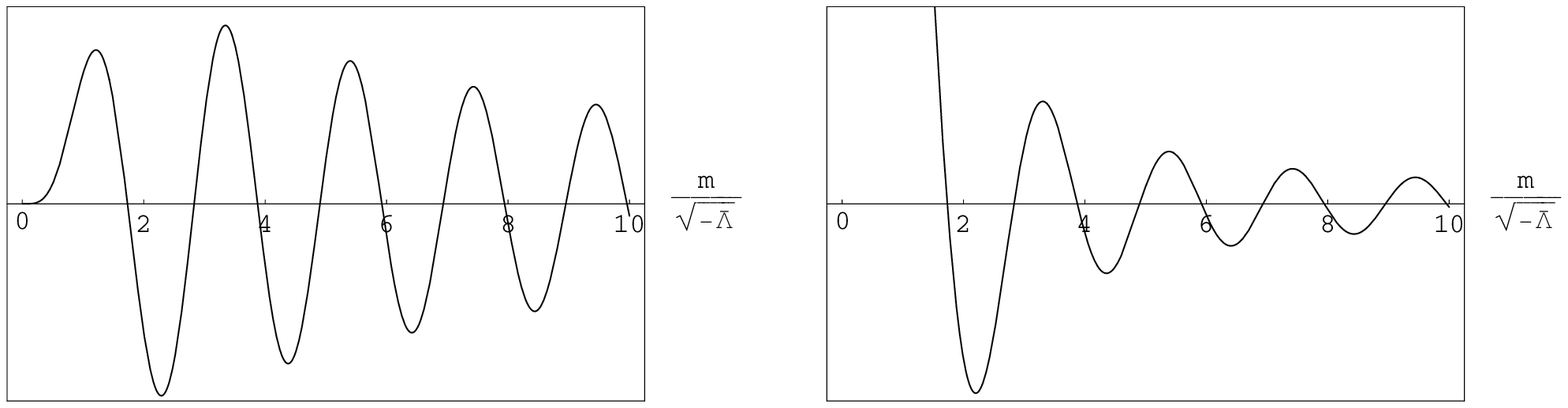, width=.95 \linewidth}             
        \caption{\small{The zeros correspond to the mass spectrum of the gravitini $\hat{\psi}^+$, $\hat{\psi}^-$, ($kr_{c}\pi=10$).} \label{masygravitina10}}
        \end{center}    
\end{figure}

The situation becomes interesting when we consider large extra dimension. In such a case masses are quantized in the unit of $\sqrt{-\bar{\Lambda}}$ (see figure \ref{masygravitonu10} and \ref{masygravitina10}). Effectively, gravitini spectra appear to be identical, with the only difference that the  tower of the $m^+$ starts one unit of  $\sqrt{-\bar{\Lambda}}$ below $m^-$.  
        Spectra of gravitons and gravitini $\hat{\psi}^+$  contain ultra-light modes. One can compute there the  masses in the limit $k r_c \gg 1$: 
	\begin{equation} m_{2}=\sqrt{12}e^{-\frac{k\pi r_{c}}{2}}\sqrt{-\bar{\Lambda}}\ ,\qquad m_{3/2^{+}}=\sqrt{8}e^{-\frac{k\pi r_{c}}{2}}\sqrt{-\bar{\Lambda}}\ . 
	 \end{equation} 
	The lightest mode of the $\hat{\psi}^-$ is much  heavier: $m_{3/2^{-}}=\sqrt{3}\sqrt{-\bar{\Lambda}}$. One can obtain this result from the general formula (\ref{spectads}) taking $E_{0}=5/2+4e^{-k\pi r_{c}}$. Note that  $4e^{-k\pi r_{c}}\ll1$.

       We can check which KK modes belong to the same supermultiplet by comparing shapes of their wave functions. Let us concentrate on the supersymmetry transformation $\delta e^a_\mu=\frac{1}{2}\bar{\eta}^A\gamma^a\Psi_{\mu A}$. Decompositions $ e^a_\mu=a(y)\bar{e}_{\mu}^a+\phi_{e}(y)\bar{h}_{\mu}^a(x^{\rho})$, where $\bar{e}_{\mu}^a$ is a 4d vielbain in $AdS_{4}$  background ($\bar{g}_{\mu\nu}=\eta_{ab}\bar{e}_{\mu}^a\bar{e}_{\nu}^b$), (\ref{decompgravitini}) and (\ref{killingwector}) lead us to the following constrain:
       \begin{equation} \label{shapesusypl}
       \frac{\phi^{(n)}_h}{\cosh\left(k|y|-\frac{k\pi r_{c}}{2}\right)}=\cosh\left(\frac{k|y|}{2}-\frac{k\pi r_{c}}{4}\right)\phi_1^{+(n)}-\sinh\left(\frac{k|y|}{2}-\frac{k\pi r_{c}}{4}\right)\phi_2^{+(n)}\ ,
       \end{equation} 
       where $(n)$ denotes the level of the KK modes. Notice that we have used
the following equality which holds for fluctuations of the metric tensor and vielbain: $\delta g_{\mu\nu}=\eta_{ab}(e^a_\mu\delta e^b_\nu+ e^b_\nu\delta e^a_\mu)$. One can check that in this case equation (\ref{shapesusypl}) is satisfied at each level of the KK tower.

\section{Localization}
         We shall analyze localization of gravitini and gravitons comparing 
five--dimensional amplitudes that multiply four--dimensional effective kinetic terms in the Lagrangian. For the Einstein--Hilbert term one can write
	 \begin{equation} 
	 \frac{1}{2}\sqrt{-G}R\supset \frac{1}{8}\sqrt{-g}a^{-2}(y)\phi^{2}_h\eta^{\mu\nu}\eta^{\rho\sigma}\eta^{\gamma\delta}\nabla_\mu h_{\rho\gamma}\nabla_\nu h_{\sigma\delta}\ ,
	 \end{equation} 
	 hence the localization of the gravitons is determined by the factor $a^{-2}(y)\phi^{2}_h(y)$.
   
     For the kinetic term of the gravitino one  writes
      \begin{eqnarray} 
	 &&-\frac{1}{2}\sqrt{-G}\bar{\Psi}^A_\alpha\gamma^{\alpha\beta\gamma}D_\beta\Psi_{\gamma A}\supset -\frac{1}{2}\sqrt{-g}\left((\phi_1^+)^{2}+(\phi_2^+)^{2}\right)a(y)\bar{\psi^+}_\mu\gamma^{\mu\nu\rho}\nabla_\nu\psi^+_{\rho}\nonumber\\&&-\frac{1}{2}\sqrt{-g}\left((\phi_1^-)^{2}+(\phi_2^-)^{2}\right)a(y)\bar{\psi^-}_\mu\gamma^{\mu\nu\rho}\nabla_\nu\psi^-_{\rho}\ .
	 \end{eqnarray} 
  It is straightforward to check that the zero modes and the ultra--light modes of the graviton and the gravitino are localized on the branes while the wave functions of the remaining are spread over the bulk.       
   
 \section{Flipped bigravity}
\begin{figure}[h]
        \begin{center}
        \epsfig{file=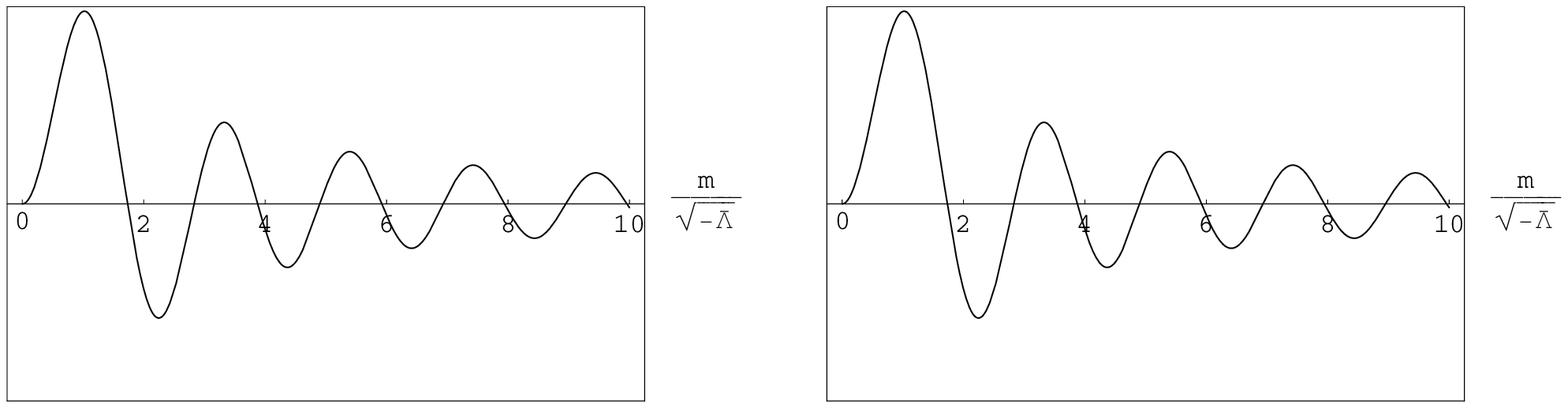, width=.95 \linewidth}
        \caption{\small{The zeros corresponds to the mass spectrum of  the gravitini $\hat{\psi}^+$, $\hat{\psi}^-$,
($k r_{c}\pi=10$).} \label{masygravitinaflip10}}  
        \end{center}
\end{figure}

 Let us return  to the flipped  bigravity, where four--dimensional vacuum spontaneously brakes all supersymmetries. To obtain such model, assume $\alpha_0=1/\alpha_\pi=-\alpha$, where
        \begin{equation} 
	\alpha=\frac{\cosh(k\pi r_{c}/2)- 1}{\sinh(k\pi r_{c}/2)}\ .
      \end{equation} 
      Then the boundary conditions assume the form 
         \begin{eqnarray} 
	&&\epsilon^{-1}(y) \delta(y) \gamma_5(\Psi^-_\mu)^A=\delta(y)\alpha(\sigma_1)^A_{\;B}(\Psi^+_\mu)^B\ ,\label{warbrzadsbigravitflip1}\\	&&\epsilon^{-1}(y) \delta(y-\pi r_c) \gamma_5(\Psi^-_\mu)^A=-\delta(y-\pi r_c)\alpha^{-1}(\sigma_1)^A_{\;B}(\Psi^+_\mu)^B\label{warbrzadsbigravitflip2}\ .
      \end{eqnarray} 
	 One can easily check that the Killing spinor does not exist. 
We factorize five--dimensional spinors as in (\ref{decompgravitini}). Quantization conditions remove zero modes from the spectra. As we can see in figure 
\ref{masygravitinaflip10}, the spectra of $\hat{\psi}^+$ and $\hat{\psi}^-$ 
are identical and they do not fit the mass formulae (\ref{spectads}). 
For a small $k r_{c}$ masses are quantized in  units of $1/r_{c}$ and the 
KK tower of the gravitini is shifted by $1/(2r_{c})$ with respect to the one
 of the graviton.  
	Notice, that even in the limit $r_{c}\rightarrow \infty$ supersymmetry is not restored. For large $r_c$ in both spectra ($\hat{\psi}^+$ and $\hat{\psi}^-$) we observe ultra-light modes with masses 
	\begin{equation} m_{3/2^{+}}= m_{3/2^{-}}=2e^{-\frac{k\pi r_{c}}{2}}\sqrt{-\bar{\Lambda}}\ . 
	 \end{equation} 

	As in the previous case, we can analyze localization of the wave functions. One can see that ultra--light modes are localized, one on each brane. 
The remaining wave functions are spread over the bulk.
	\section{Summary}
	We have constructed a class of five--dimensional supergravities 
on an orbifold $S^{1}/Z_{2}$ with localized sources at the fixed points. 
Supersymmetry  requires special relations between brane tensions $\lambda_0$, $\lambda_\pi$ and gravitini masses $\alpha_0$, $\alpha_\pi$ on each brane. 
We have performed Kaluza--Klein decomposition of 5d super--bigravities 
corresponding to 
periodic and twisted boundary condition for gravitini fields. 
In the first case, gravitational background preserves one half of 
the supersymmetries, while in the second supersymmetry is completely 
(but spontaneously) broken. In the limit $r_c \rightarrow\infty$ we have obtained ultra--light modes in the spectra of gravitons and gravitini. 
We have analyzed localization of the wave functions. In the bosonic sector massless and ultra--light modes are localized on the both branes while others are spread over the bulk. Gravitini in the unbroken case are localized in the 
same way. In the case of spontaneously broken supergravity we do not find massless modes of the gravitini, but there exist two ultra--light gravitini modes instead. Their wave functions are localized on different branes.  
The wave functions of the remaining (heavy) modes are spread over the bulk.    

\section*{Acknowledgments}   
Z.L. thanks Theory Division at CERN for hospitality.
\noindent This work  was partially supported  by the EC Contract
HPRN-CT-2000-00152 for years 2000-2004, by the Polish State Committee for Scientific Research grants KBN 2P03B 001 25 (Z.L.) and by KBN 2P03B 124 25 (R.M.).


\begin{thebibliography}{99} 

\bibitem{Altendorfer:2000rr}
R.~Altendorfer, J.~Bagger and D.~Nemeschansky,
Phys.\ Rev.\ D {\bf 63} (2001) 125025

\bibitem{Falkowski:2000er}
A.~Falkowski, Z.~Lalak and S.~Pokorski,
Phys.\ Lett.\ B {\bf 491} (2000) 172

\bibitem{Lalak:2003fu}
Z.~Lalak and R.~Matyszkiewicz,
Phys.\ Lett.\ B {\bf 562} (2003) 347

\bibitem{Bergshoeff:2000zn}
E.~Bergshoeff, R.~Kallosh and A.~Van Proeyen,
JHEP {\bf 0010} (2000) 033

\bibitem{Brax:2001xf}
P.~Brax, A.~Falkowski and Z.~Lalak,
Phys.\ Lett.\ B {\bf 521} (2001) 105

\bibitem{Bagger:2002rw}
J.~Bagger and D.~V.~Belyaev,
Phys.\ Rev.\ D {\bf 67} (2003) 025004

\bibitem{Kogan:1999wc}
I.~I.~Kogan, S.~Mouslopoulos, A.~Papazoglou, G.~G.~Ross and J.~Santiago,
Nucl.\ Phys.\ B {\bf 584} (2000) 313

\bibitem{Kogan:2000vb}
I.~I.~Kogan, S.~Mouslopoulos and A.~Papazoglou,
Phys.\ Lett.\ B {\bf 501} (2001) 140

\bibitem{Lalak:2002kx}
Z.~Lalak and R.~Matyszkiewicz,
Nucl.\ Phys.\ B {\bf 649} (2003) 389

\end{thebibliography}
\end{document}